\documentclass[a4paper]{jpconf}
\begin{document}
\title{Deterministic models of quantum fields}

\author{Hans-Thomas Elze}

\address{Dipartimento di Fisica, Via Filippo Buonarroti 2, I-56127 Pisa, Italia}

\ead{Elze@df.unipi.it}

\begin{abstract}
Deterministic dynamical models are discussed which can be described in quantum mechanical terms.  
-- In particular, a local {\it quantum} field theory is presented which {\it is} a supersymmetric 
{\it classical} model \cite{I05}. The Hilbert space approach of Koopman and von\,Neumann is used to study the classical evolution of an ensemble of such systems. Its Liouville operator is decomposed into two contributions, with positive and negative spectrum, respectively. The unstable negative part is eliminated by a constraint on physical states, which is invariant under the Hamiltonian flow. Thus, choosing suitable variables, the classical Liouville equation becomes a functional Schr\"odinger equation of a genuine quantum field theory. 
-- We briefly mention an U(1) gauge theory with ``varying alpha'' or dilaton coupling where a corresponding quantized theory emerges in the phase space approach \cite{I05U1}. 
It is energy-parity symmetric and, therefore, a prototype of a model in which the cosmological constant is protected by a symmetry.  
\end{abstract}

\section{Introduction}
The (dis)similarity between the classical Liouville 
equation and the Schr\"odinger equation has recently been discussed, considering 
this an appropriate starting point for attempts to derive quantum from classical dynamics, 
i.e. for {\it emergent quantum theory} \cite{I05,I05U1}. 

In suitable coordinates both equations
are remarkably similar, apart from the characteristic doubling of the classical 
phase space degrees of freedom as compared to the quantum mechanical case. The Liouville 
operator is Hermitian in the operator approach to classical statistical mechanics 
developed by Koopman and von\,Neumann \cite{KN}. However, unlike the case of the 
quantum mechanical Hamiltonian, its spectrum is generally not bounded from below.  
Therefore, attempts to find a deterministic foundation of quantum theory must pay special attention to the construction of a stable ground state, if they are based on a classical 
ensemble theory.   
 
Research in this direction is suggested by work of 
't\,Hooft, demonstrating several examples of systems 
which can be faithfully described as quantum mechanical and yet  
present deterministic dynamical models.
It has been argued in favour of such model building that it may 
lead to a new approach in trying to understand and possibly resolve 
the persistent clash between general relativity and quantum theory, 
by questioning the fundamental character of the latter 
\cite{tHooft01}. 

There have always been speculations 
about the (im)possibility of deriving quantum theory from more fundamental and deterministic 
dynamical structures. The discourse running from Einstein, Podolsky and Rosen 
to Bell, with numerous successors, is well known.
Much of this has come under experimental scrutiny and   
no disagreement with quantum theory has been 
observed in the laboratory experiments on scales very large compared to the Planck scale.   
Nevertheless, it is conceivable that quantum mechanics 
emerges only on sufficiently large scales, where it describes effectively the fundamental deterministic degrees of freedom.  

A class of particularly simple emergent quantum models comprises systems which classically 
evolve in discrete time steps \cite{tHooft01,ES02}. Pointing towards a 
general nonlocality feature is the finding here that the coordinate eigenstates of the emergent quantum system are related to superpositions of underlying ``primordial'' states, which refer to the position of the classical degree of freedom. -- Employing the path integral formulation of 
classical mechanics introduced by Gozzi and collaborators \cite{Gozzi}, it has been shown that 
classical models of Hamiltonian dynamics similarly turn into unitary quantum mechanical ones, if the corresponding Liouville operator governing the evolution of phase space densities is 
discretized \cite{I04}. Here, the arbitrariness in such discretizations helps to find a stable groundstate. Models of intrinsically discrete nature should be interesting to study in this context, such as causal sets.     

Furthermore, it has been observed that classical systems with Hamiltonians which are linear 
in the momenta, can generally be represented in quantum mechanical terms. 
However, a new kind of gauge fixing or constraints implementing 
``information loss'' at a fundamental level have to be invoked, in order 
to provide a groundstate for such systems    
\cite{tHooft01,Vitiello01,Blasone04}. -- 
Various other arguments for deterministically induced quantum features have been proposed 
recently;  
see works collected in Part\,III of Ref.\,\cite{E04}, for example, or Refs.\,\cite{Smolin,Adler}, 
concerning statistical and/or dissipative systems, quantum gravity, and matrix models. 
In all cases, the unifying dynamical 
principle leading to the necessary truncation of the Hilbert space is still  
missing.   

Presently, I present a    
deterministic field theory from which a corresponding {\it quantum theory     
emerges by constraining the classical dynamics}. A functional Schr\"odinger equation is obtained with a positive Hamilton operator, which involves a standard scalar boson part. Main ingredient is a splitting of the 
Liouville operator into positive and negative energy contributions. 
The latter would render the to-be-quantum field theory unstable and are   
eliminated by a constraint on the physical states, based on   
the supersymmetry of the classical system. 
This is analogous to the ``loss of information'' 
condition in 't\,Hooft's and subsequent work \cite{tHooft01,Vitiello01,Blasone04}. We   
hope that the study of interacting fields will lead to    
better understand the dynamical origin of such a constraint. A   
dissipative information loss mechanism is plausible, yet a  
dynamical symmetry breaking may be an alternative \cite{I05U1}. 

It seems worth while to 
point out still another perspective on the emergence of 
quantum mechanics from possibly more fundamental physics.  
Textbooks explain {\it how to 
quantize} a given classical system, following the rules of imposing commutators or of setting up 
a Feynman path integral, etc. These 
reflect the status of the experimentally acquired knowledge. However, where do   
the quantization rules come from? -- 
This question surfaces now and then since the early days of quantum theory. 
It must not be forgotten that quantum theory is beset with serious 
problems, other than the incompatibility with general relativity. -- 
The infinities of quantum field theory are dealt with by  
renormalization. This procedure has become familiar to the extent of not 
perceiving it as problematic anymore, even if it may rule out a theoretical determination of   
its basic parameters (masses, couplings, etc.). -- The emergence of the classical 
world of our experience from the quantum mechanical picture was a problem that has been solved.  
It is understood through {\it environment induced decoherence}, i.e., as 
being due to the interaction of quantum mechanical systems with 
the ``rest of the universe'' \cite{ZEHetal,Zurek,Omnes}. However, related is the 
measurement problem which states that a measurement on a 
quantum system which leads to a classical apparatus reading is a process which 
cannot be described entirely and consistently within quantum theory itself 
\cite{Adler,Bastin,WheelerZurek}. This problem has not been solved. 
It has given rise to a number 
of dynamical wave function collapse or reduction models, however, 
with no generally accepted completion of quantum physics in this respect    
\cite{AdlerM,Diosi}. -- 
Clearly, there is a need to better understand 
or change the foundations of quantum theory.    

\section{A quantum field as a supersymmetric classical one}
We will make use of ``pseudoclassical mechanics'' or, rather,  
pseudoclassical field theory \cite{I05}. These notions have been 
introduced through the work of Refs.\,\cite{CB}, considering 
a {\it Grassmann variant of classical mechanics} and  
studying the dynamics of spin degrees of freedom classically (and 
after quantization in the usual way). -- Consider a ``fermionic'' field $\psi$, together with a real scalar field $\phi$. 
The former is represented by the nilpotent generators of an infinite dimensional 
Grassmann algebra. They obey:  
\begin{equation}\label{odd} 
\{ \psi (x),\psi (x')\}_+\equiv\psi (x)\psi (x')+\psi (x')\psi (x)=0 
\;\;, \end{equation}
where $x,x'$ are coordinate labels in Minkowski space. All elements are real. 
Then, the classical model to be studied is defined by the action: 
\begin{equation}\label{action}
S\equiv\int\mbox{d}^4x\;\Big (\dot\phi\dot\psi -\phi\big (-\Delta +m^2+v(\phi)\big )\psi\Big )\equiv\int\mbox{d}t\;L 
\;\;, \end{equation} 
where dots denote time derivatives, and $v(\phi )$ may be a polynomial in $\phi$, for example. 
Note that the action is Grassmann odd; such kind of 
models have been studied in different context by Volkov et al. 
   
In terms of canonical momenta, $P_\phi\equiv\delta L/\delta\dot\phi =\dot\psi$,  
$P_\psi\equiv\delta L/\delta\dot\psi =\dot\phi$,  
the Hamiltonian is:  
\begin{equation}\label{Hamiltonian} 
H=\int\mbox{d}^3x\;\Big (P_\phi\dot\phi+P_\psi\dot\psi\Big )-L 
=\int\mbox{d}^3x\;\Big (P_\phi P_\psi +\phi K\psi\Big )  
\;\;, \end{equation} 
where $K\equiv -\Delta +m^2+v(\phi )$ and, for later use, 
$K'\equiv K+\phi\mbox{d}v(\phi )/\mbox{d}\phi$. Then,    
Hamilton's equations can be shown to be invariant under 
two global symmetry transformations, 
\begin{equation}\label{sym1} 
\phi\longrightarrow\phi +\epsilon\psi\;\;;\;\;\;
\psi\longrightarrow\psi +\epsilon\dot\phi      
\;\;, \end{equation} 
where $\epsilon$ is an infinitesimal real parameter. Associated are the conserved 
Noether charges:  
\begin{equation}\label{C1} 
C_1\equiv\int\mbox{d}^3x\;P_\phi\psi\;\;;\;\;\;
C_2\equiv\int\mbox{d}^3x\;\Big (\frac{1}{2}P_\psi^2+V(\phi )\Big )   
\;\;. \end{equation} 
The second one is the total energy of the scalar 
field, with $\mbox{d}V(\phi )/\mbox{d}\phi\equiv K\phi$.  

We introduce the Poisson bracket of   
observables $A$,$B$ defined over phase space: 
\begin{equation}\label{PB} 
\{A,B\}\equiv A\int\mbox{d}^3x\;\Big (     
\frac{\stackrel{\leftharpoonup}{\delta}}{\delta P_\phi}\frac{\stackrel{\rightharpoonup}{\delta}}{\delta\phi}
+\frac{\stackrel{\leftharpoonup}{\delta}}{\delta P_\psi}\frac{\stackrel{\rightharpoonup}{\delta}}{\delta\psi}
-\frac{\stackrel{\leftharpoonup}{\delta}}{\delta\phi}\frac{\stackrel{\rightharpoonup}{\delta}}{\delta P_\phi}
-\frac{\stackrel{\leftharpoonup}{\delta}}{\delta\psi}\frac{\stackrel{\rightharpoonup}{\delta}}{\delta P_\psi}
\Big )B  
\;\;. \end{equation} 
Functional derivatives refer to the same space-time point and act in the 
indicated direction; it coincides with their fermionic left/right-derivative character. 
Embodying Hamilton's equations of motion, this yields the familiar relation: 
\begin{equation}\label{timederivs} 
\frac{\mbox{d}}{\mbox{d}t}A=\{ H,A\}+\partial_tA
\;\;. \end{equation}
For the Hamiltonian and Noether charge densities, identified by  
$H\equiv\int\mbox{d}^3xH(x)$ and $C_j\equiv\int\mbox{d}^3xC_j(x)|_{j=1,2}$, 
respectively, one finds a local (equal-time) supersymmetry algebra, 
see the second of Refs.\,\cite{I05}.   
A Hilbert space version of the symmetry algebra will be obtained  
shortly. -- An important example of Eq.\,(\ref{timederivs}) is the Liouville equation.  
Considering an ensemble of systems, this equation governs the evolution of its phase 
space density $\rho$: 
\begin{equation}\label{Liouville} 
0=i\frac{\mbox{d}}{\mbox{d}t}\rho =i\partial_t\rho -\hat{\cal L}\rho
\;\;, \end{equation}
where a convenient factor $i$ has been introduced, and the Liouville operator 
$\hat{\cal L}$ is defined by: 
\begin{equation}\label{Liouvilleop}  
-\hat{\cal L}\rho\equiv i\{ H,\rho \}
\;\;. \end{equation}
These equations summarize the classical statistical mechanics of a conservative system.  

An equivalent Hilbert space formulation is due to
Koopman and von\,Neumann \cite{KN}. It will be modified appropriately 
for our supersymmetric classical field theory.  
Two {\it postulates} are put forth: 
({\bf A}) the phase space density functional can be factorized in the form $\rho\equiv\Psi^*\Psi$;
({\bf B}) the Grassmann valued and, in general, complex state functional $\Psi$ itself obeys the 
Liouville Eq.\,(\ref{Liouville}). --
Furthermore, the complex valued inner product of such state functionals is defined by: 
\begin{equation}\label{scalarprod}
\langle\Psi |\Phi\rangle\equiv\int {\cal D}\phi {\cal D}P_\psi {\cal D}\psi {\cal D}P_\phi\; 
\Psi^*\Phi =\langle\Phi |\Psi\rangle^*
\;\;, \end{equation}
i.e., by functional integration over all phase space variables. Since there are  
Grassmann valued variables, the $*$-operation defining   
the dual $\Psi^*$ needs to be treated carefully. -- 
Given the Hilbert space structure, the Liouville operator of a conservative system is  
Hermitian and the overlap $\langle\Psi |\Psi\rangle$ is a conserved quantity. 
Then, the Liouville equation also applies to 
$\rho =|\Psi |^2$, due to its linearity, and $\rho$ may be  
interpreted as a phase space density, as before \cite{KN}. 
    
Certainly, one is reminded here of quantum mechanics. -- In order to 
expose the striking similarity as well as the remaining crucial difference, further transformations of the functional Liouville equation are useful \cite{I05,I05U1}.  
Fourier transformation replaces the momentum $P_\psi$ by a second scalar field      
$\bar\phi$. Furthermore, define $\bar\psi\equiv P_\phi$.  
Thus, the Eqs.\,(\ref{Liouville})--(\ref{Liouvilleop}) yield: 
\begin{equation}\label{Schroedinger} 
i\partial_t\Psi =\hat{\cal H}\Psi
\;\;, \end{equation} 
where $\Psi$ is considered as a functional of $\phi,\bar\phi,\psi,\bar\psi$, 
and with the {\it emergent} ``Hamilton operator'': 
\begin{equation}\label{Hem}
\hat{\cal H}\Psi\equiv -i\int {\cal D}P_\psi \;\exp (iP_\psi\cdot\bar\phi )\{ H,\Psi\} 
=\int\mbox{d}^3x\;\Big (-\delta_{\bar\phi}\delta_\phi 
+\bar\phi K\phi 
-i(\bar\psi\delta_\psi -\psi K'\delta_{\bar\psi})\Big )\Psi
\;\;, \end{equation}
using $f\cdot g\equiv\int\mbox{d}^3x\;f(x)g(x)$. The Hamiltonian (density) 
is Grassmann even. -- 
While Eq.\,(\ref{Schroedinger}) appears as a {\it functional Schr\"odinger equation}, 
several remarks are in order here. First, following the transformation, 
$\phi\equiv (\sigma +\kappa )/\sqrt 2$ and 
$\bar\phi\equiv (\sigma -\kappa )/\sqrt 2$, one finds a ``bosonic''  
kinetic energy: 
\begin{equation}\nonumber 
-\frac{1}{2}\int\mbox{d}^3x\;\left (\delta_\sigma^{\;2}-\delta_\kappa^{\;2}\right ) 
\;\;, \end{equation} 
which is {\it not bounded from below}. This means that the Hamiltonian lacks a lowest energy state. Secondly, the $*$-operation mentioned before amounts to complex conjugation for a bosonic 
state functional, in analogy with     
a quantum mechanical wave function. However, based on complex conjugation alone, 
the fermionic part of the Hamiltonian (\ref{Hem}) would not be Hermitian. 
Instead, a detailed construction of the inner 
product for functionals of Grassmann valued fields has been given and applied, respectively,  
in Refs. \cite{Jackiw} and \cite{I05,I05U1}. 
Considering the {\it noninteracting case} 
with $K'=K$, i.e., with $v(\phi )=0$ in Eq.\,(\ref{action}), the construction of   
Floreanini and Jackiw suffices here: the Hermitian 
conjugate of $\psi$ is $\psi^\dagger =\delta_\psi$ and of $\bar\psi$ it is $\bar\psi^\dagger =\delta_{\bar\psi}$. 
Furthermore, rescaling $\bar\psi\;\longrightarrow\;\bar\psi\sqrt K$,    
the Hermitian fermionic part of the Hamiltonian (\ref{Hem}) becomes: 
\begin{equation}\label{HemF} 
\hat{\cal H}_{\bar\psi\psi}\equiv i(\psi\sqrt K\delta_{\bar\psi} -\bar\psi\sqrt K\delta_\psi )
\;\;. \end{equation}  
In the presence of interactions, with $K'\neq K$, additional modifications are 
necessary, which will not be considered here. 
In any case, the eigenvalues of $\hat{\cal H}_{\bar\psi\psi}$ 
will not have a lower bound either. 

To summarize, the emergent Hamiltonian $\hat{\cal H}$ 
is unbounded from below, lacking a groundstate. 
This difficulty has been encountered in various other attempts to build deterministic 
quantum models \cite{I05U1,tHooft01,ES02,I04,Vitiello01,Blasone04}. For our case, 
it will be solved next.

\section{Groundstate construction for the emergent quantum model}
We proceed with equal-time operator relations for the interacting case, 
which are related to the supersymmetry algebra mentioned before \cite{I05}. 
Using Eqs.\,(\ref{C1}) and $\bar\psi\equiv P_\phi$, as before, 
one obtains:  
\begin{equation}\label{C1op} 
\hat{\cal C}_1(x)\Psi\equiv\int {\cal D}P_\psi \;\exp (iP_\psi\cdot\bar\phi )\{ C_1(x),\Psi\}
=\big (-\psi\delta_\phi +i\bar\psi\bar\phi\big )_{(x)}\Psi 
\;\;. \end{equation} 
Analogously, we calculate: $\hat{\cal C}_2(x)\Psi\;\equiv\;
\big (-i\delta_{\bar\phi}\delta_\psi -\phi K\delta_{\bar\psi}\big )_{(x)}\Psi$.  
Both operators are Grassmann odd and obey: 
$\{ \hat{\cal C}_j(x),\hat{\cal C}_j(x')\}_+=0$, for $j=1,2$. 
Furthermore, one finds the vanishing commutator: 
\begin{equation}\label{Hcomm} 
[\hat{\cal H}(x),\hat{\cal H}(x')]=0 
\;\;, \end{equation} 
i.e., the emergent theory is {\it local}. 
However, the Hamilton operator, Eq.\,(\ref{Hem}), involves a functional Fourier transform. Therefore, the emergent quantum field theory, which is local in the usual sense, is {\it nonlocal} with respect to the space of fields of the underlying classical system. This is analogous to what has been found in several models mentioned in the Introduction \cite{tHooft01,ES02,Smolin}; see also the third of Refs.\,\cite{I05} and \cite{I05U1} for more detailed discussions. -- Finally, 
we find: 
\begin{equation}\label{SUSYop1} 
[\hat{\cal H}(x),\hat{\cal C}_j(x')]=0\;\;,\;\;\mbox{for}\;j=1,2 
\;\;;\;\;\;
\{ i\hat{\cal C}_1(x),\hat{\cal C}_2(x')\}_+=\hat{\cal H}(x)\delta^3(x-x')
\;\;. \end{equation} 
One may complete these with relations for the full set of  
operators generating the space-time symmetries of our model. 
However, they do not play a special role in the following.  

Since the emergent Hamiltonian 
lacks a proper groundstate, we cannot yet interpret the model as a quantum mechanical one, 
despite close formal similarities. -- This problem is solved, if we  
find a positive definite local operator $\hat P$ that obeys  
$[\hat{\cal H}(x),\hat P(x')]=0$. Then, the Hamiltonian can be split into 
contributions with positive and negative spectrum: 
\begin{equation}\label{Hemsplit} 
\hat{\cal H}=\hat{\cal H}_+-\hat{\cal H}_-
\;\;, \end{equation}
where, in the simplest way,  
$\hat{\cal H}_\pm (x)\equiv (\hat{\cal H}(x)\pm\hat P(x))^2/4\hat P(x)$. 
With this, the spectrum of the Hamiltonian $\hat{\cal H}$ is made 
bounded from below by imposing the ``positivity constraint'': 
\begin{equation}\label{constraint} 
\hat{\cal H}_-\Psi =0 
\;\;, \end{equation}
which is conserved, since $[\hat{\cal H}_+(x),\hat{\cal H}_-(x)]=0$,  
by construction. In this way, the {\it physical states} are selected  
which are based on the existence of a {\it quantum mechanical groundstate}. 

For our field theory, the noninteracting and interacting cases   
have been studied separately in the second of Refs.\,\cite{I05}. 
We specialize here to the noninteracting case. As mentioned before, with $v(\phi )=0$ in Eq.\,(\ref{action}), and therefore $K'=K=-\Delta +m^2$, the rescaling $\bar\psi\;\longrightarrow\;\bar\psi\sqrt K$ is useful, and we consider the operators: 
$\hat{\cal H}(x)=\big (-\delta_{\bar\phi}\delta_\phi +\bar\phi K\phi \big )_{(x)}
+\hat{\cal H}_{\bar\psi\psi}(x)$, 
$i\hat{\cal C}_1(x)=
\big (-i\psi\delta_\phi -\bar\psi\sqrt K \bar\phi\big )_{(x)}$, 
$\hat{\cal C}_2(x)=
\big (-i\delta_{\bar\phi}\delta_\psi -\phi\sqrt K \delta_{\bar\psi}\big )_{(x)}$, 
with $\hat{\cal H}_{\bar\psi\psi}$ from Eq.\,(\ref{HemF}). 
Completing the $\hat{\cal C}_j$ to Hermitian operators, with  
$\psi^\dagger =\delta_\psi$ and 
$\bar\psi^\dagger =\delta_{\bar\psi}$, a suitable combination 
yields an operator $\hat P$ with the required properties; in terms of   
{\it ``square-root of harmonic oscillator''} operators,  
$\hat{\cal C}_{1+}(x)\;\equiv\;\big (i\hat{\cal C}_1(x)+\big (i\hat{\cal C}_1(x)\big )^\dagger\big )$ and $\hat{\cal C}_{2+}(x)\;\equiv\;\big (\hat{\cal C}_2(x)+\big (\hat{\cal C}_2(x)\big )^\dagger\big )$:  
$\hat P(x)\equiv\big (\hat{\cal C}_{1+}^{\;2}(x)+\hat{\cal C}_{2+}^{\;2}(x)\big )/\delta^3(0)$.  
This results in: 
$\hat{\cal H}_\pm (x)
=\big (-\delta_\phi^{\;2}+\phi K\phi -\delta_{\bar\phi}^{\;2}+\bar\phi K\bar\phi\big )/4  
\pm\hat{\cal H}(x)/2+\hat{\cal H}^2(x)/4\hat P(x)$, 
cf. Eq.\,(\ref{Hemsplit}). Finally, with $\phi\equiv (\sigma +\kappa )/\sqrt 2$ and 
$\bar\phi\equiv (\sigma -\kappa )/\sqrt 2$, the Hamiltonian density becomes: 
\begin{equation}\label{Hemplus2}
\hat{\cal H}_+(x)=\frac{1}{2}\Big (-\delta_\sigma^{\;2}+\sigma K\sigma 
+\hat{\cal H}_{\bar\psi\psi}+\frac{1}{2}\hat{\cal H}^2/\hat P\Big )_{(x)} 
\;\;. \end{equation} 
The only trace of the previous instability is now relegated to the last   
term. Similarly, the constraint operator becomes: 
$\hat{\cal H}_-(x)=\big (-\delta_\kappa^{\;2}+\kappa K\kappa  
-\hat{\cal H}_{\bar\psi\psi}+\hat{\cal H}^2/2\hat P\big )_{(x)}/2$.  
Note the symmetry with $\hat{\cal H}_+$  
($-\hat{\cal H}_{\bar\psi\psi}=\hat{\cal H}_{\psi\bar\psi}$). Thus, elimination 
of part of the Hilbert space, Eq.\,(\ref{constraint}), may be related to a symmetry 
breaking \cite{I05}. -- The resulting Hamilton operator $\hat{\cal H}_+$ now 
has a positive spectrum, by construction, and the leading terms 
are those of a {\it free bosonic quantum field} together with a {\it fermion doublet}  
in the Schr\"odinger representation. They dominate at low energy.   

A different solution of the lacking groundstate problem has been 
proposed in Ref.\,\cite{I05U1}: It is shown that energy-parity arises in the 
Hilbert space representation 
of {\it classical phase space dynamics} of matter. (This symmetry was introduced 
to protect the cosmological 
constant against large matter contributions \cite{KS05}.)
Generalizing an Abelian gauge theory by a varying coupling, as in ``varying alpha'' or 
dilaton models, classical matter fields can turn 
into {\it quantum fields} (Schr\"odinger picture), accompanied by a gauge symmetry change  
-- U(1)$\;\rightarrow\;$U(1)\,x\,U(1). The transition between classical ensemble theory and quantum field theory is governed here by the varying coupling through correction terms that 
introduce diffusion and dissipation. 
  
\section{Conclusions}
We discussed {\it deterministic (ensemble) theories} trying to reconstruct  
quantum mechanics as an emergent phenomenon, which may  
challenge common wisdom about its foundations and limitations . -- 
More immediate consequences for the 
measurement process, reduction, and the ``collapse of the wave function'' \cite{ZEHetal,Omnes,Bastin,WheelerZurek} need to be explored, as well as extensions  
to more realistic theories, such as the Standard Model including gravity. 


\end {document}